\documentclass[prd,aps,preprint,amsmath,amssymb,eshowkeys,nofootinbib]{revtex4-2}
\usepackage[mathscr]{euscript}
\usepackage{dcolumn}
\usepackage{multirow}
\usepackage{bm}
\usepackage{graphicx}
\usepackage{subfigure}
\usepackage{rotating}
\usepackage{amsmath,amssymb,amsthm}
\usepackage{changepage}
\usepackage[colorlinks=true,linkcolor=red,urlcolor=green,citecolor=green]{hyperref}
\newcommand{\bea}{\begin{eqnarray}}
\newcommand{\eea}{\end{eqnarray}}
\newcommand{\beq}{\begin{equation}}
\newcommand{\eeq}{\end{equation}}

\def\/{\over}

\begin{document}

\title{Analytical primordial power spectra from a potential dominated pre-inflationary bounce in a closed universe}
\author{Qihong Huang\footnote{Corresponding author: huangqihongzynu@163.com}}

\affiliation{
School of Physics and Electronic Science, Zunyi Normal University, Zunyi, Guizhou 563006, China
}

\begin{abstract}
The CMB TT spectrum is suppressed at large scales, a feature that may originate from pre-inflationary dynamics in a closed universe. We study the primordial power spectrum for a potential dominated pre-inflationary bounce in a spatially closed universe by extending the analytical approximations of Thavanesan et al.~\cite{Thavanesan2021}, and find that it exhibits the same suppression trend as the KICI and emergent scenarios. Using the analytical approximation, we further show that the resulting CMB TT spectra for the pre-inflationary bounce scenario are the same as those of the KICI and emergent scenarios, indicating that the pre-inflationary bounce scenario can account for the large scale suppression of the CMB TT spectrum as well as the KICI and emergent scenarios. This extends the potential independent analytical approach to a potential dominated pre-inflationary bounce, providing a new scenario that accounts for the observed large scale suppression of the CMB TT spectrum.
\end{abstract}

\maketitle

\section{Introduction}

Inflation is an epoch of exponential expansion in the early universe, which resolves the horizon and flatness problems of the standard cosmological model~\cite{Guth1981, Linde1982}. The scalar perturbations produced during inflation, which are adiabatic, Gaussian, and nearly scale-invariant, provide the seed for both the Cosmic Microwave Background (CMB) temperature anisotropies and the Large-Scale Structure of the universe~\cite{Mukhanov1981, Lewis2000, Bernardeau2002}, the latter commonly characterized by the matter power spectrum. These predictions have been confirmed by observations of the CMB radiation from COBE~\cite{Smoot1992}, WMAP~\cite{Hinshaw2013}, and Planck~\cite{Planck2020}, and these observations also indicate a suppression of the CMB TT spectrum at large scales, which acts as a powerful probe of the physics of the very early universe, including the inflationary models, the initial conditions and the geometry of the early universe. Nevertheless, the standard slow-roll inflation model fails to explain this observational result, as it predicts a nearly-invariant primordial power spectrum. 

To generate the observed large scale suppression in the CMB TT spectrum, researchers have proposed various mechanisms to generate the observed suppression. One approach is the introduction of a cutoff in the primordial power spectrum, which features an infrared cutoff on the horizon scale as its most prominent characteristic~\cite{Bridle2003, Cline2003}. Another approach involves modifying the inflationary potential itself, which typically requires fine-tuning to produce either a sharp step-like feature~\cite{Gonzalez2019} or an oscillatory component~\cite{Kamerkar2023}. Furthermore, specific inflationary models have been constructed to generate the observed suppression, including the kinetic initial conditions for inflation (KICI) scenario~\cite{Contaldi2003, Handley2014, Hergt2019a, Hergt2019b, Thavanesan2021, Shumaylov2022}, the bouncing universe~\cite{Piao2004, Liu2013, Cai2018, Agullo2021}, and the emergent universe~\cite{Labrana2015, Huang2022, Huang2023a, Huang2023b}, with the common aim of suppressing large-scale power spectrum through alterations to the pre-inflationary evolution history. 

Recently, the KICI scenario has been extended to curved spacetime by incorporating spatial curvature~\cite{Thavanesan2021}. In this model, analytical approximations for the primordial power spectrum in the curved spacetime are obtained by considering a kinetically dominated stage before ultra-slow roll inflation, leading to suppressed primordial power spectrum and a correspondingly suppressed CMB TT spectrum. To realize this transition, an instantaneous transition is assumed between the kinetically dominated stage and the ultra-slow-roll inflation. This approach, known as the Contaldi approximation, provides a method for generating the primordial power spectrum that is independent of inflationary potential~\cite{Contaldi2003, Thavanesan2021}. However, recent research has shown that the instantaneous transition in the Contaldi approximation is unsmooth, as it corresponds to an implicit inflationary potential of the Heaviside step function form~\cite{Dineen2024}.

When the stage preceding inflation is taken to be the Einstein static state in the closed universe, the emergent scenario is proposed~\cite{Ellis2004a, Ellis2004b}, which is intended to resolve the big bang singularity~\cite{Campo2007, Wu2010, Cai2012, Zhang2014, HuangQ2015, Shabani2017, Shabani2019, HuangQ2020, Khodadi2016, Heydarzade2016, Khodadi2018, Labrana2019, Li2019, Bengochea2021, Ilyas2021, Khodadi2022}. It can also produce the observed suppression of the CMB TT spectrum through the approach proposed in~\cite{Thavanesan2021}, which is independent of the form of the potential~\cite{Huang2022, Huang2023a, Huang2023b}. Unlike the KICI scenario, the emergent scenario is dominated by both the kinetic and potential terms before inflation~\cite{Ellis2004a}, and its realization requires a smooth and explicitly constructed potential~\cite{Ellis2004b, Zhang2014, HuangQ2020}.

In the KICI scenario, the pre-inflationary stage is kinetic dominated, whereas in the emergent scenario it is dominated by both the kinetic and potential terms. However, the case with the pre-inflationary stage dominated by potential has not been analyzed. Motivated by this, we adopt the analytical approach proposed in~\cite{Thavanesan2021} to study a potential dominated pre-inflationary stage in a closed universe. We find that this stage undergoes a bounce, corresponding to a pre-inflationary bounce scenario, in which a contracting phase precedes inflation~\cite{Qiu2026, Choudhury2024, Vicente2023, Cai2018, Falciano2008, Lilley2011, Mielczarek2008, Xia2014, Liu2014, Qiu2015, Wan2015, Li2016, Ni2018}. Unlike a general bouncing universe~\cite{Gasperini1993, Molina-Paris1999, Khoury2001, Peter2002, Finelli2002, Xue2010, Ijjas2016, Gielen2016, Date2005, Ganguly2019, Tukhashvili2023, Novello2008, Singh2006, Laguna2007, Mielczarek2008, Cianfrani2010, Wilson-Ewing2013, Amoros2013, Odintsov2014, Haro2014, Cai2014, Haro2015, Haro2017, Haro2018, Martin-Benito2021, Li2021, Barragan2009, Bamba2014a, Paul2014, Bhattacharya2016, Ilyas2021, Banerjee2022}, here the bounce is followed by an inflationary stage. Since the potential term dominates throughout the whole evolution, the potential remains smooth, and no instantaneous transition is required. We obtain the analytical approximations for the primordial power spectra, which predict a large scale suppression of the CMB TT spectrum independently of the specific form of the potential. This extends the pre-inflationary bounce scenario to the potential-dominated case in a closed universe, which is complementary to the KICI and emergent scenarios.

The main goal of this paper is to study the potential dominated pre-inflationary stage in a closed universe and its imprints on the primordial power spectrum and the CMB TT spectrum. The paper is organized as follows. In Section II, we study the potential dominated pre-inflationary stage in a closed universe. In Section III, we calculate the primordial power spectrum and the CMB TT spectrum for this model. Finally, our main conclusions are presented in Section IV.

\section{Pre-inflationary bounce}

To analyze the evolution of the universe in its early stages, we consider a homogeneous and isotropic universe described by the Friedman--Lema$\hat{i}$tre--Robertson--Walker (FLRW) spacetime with the line-element
\beq
ds^2=a(\eta)^2 d\eta^2-a(\eta)^2 \bigg[\frac{dr^2}{1-K r^2}+r^2(d\theta^2+\sin^2\theta d\phi^2)\bigg],
\eeq
where $\eta$ is conformal time, $a(\eta)$ is the conformal scale factor, and $K=0,1,-1$ represent a spatially flat, closed, or open universe, respectively. The Friedmann and Klein–Gordon equations in the spatially closed universe then take the form
\beq\label{CB1}
\mathcal{H}^{2}=\frac{\kappa^{2}}{3}\Big(\frac{1}{2}\phi'^{2}+a^{2}V\Big)-1,
\eeq
\beq\label{CB2}
2\mathcal{H}'+\mathcal{H}^{2}=-\kappa^{2} \Big(\frac{1}{2}\phi'^{2}-a^{2}V\Big)-1,
\eeq
\beq\label{CB3}
\phi''+2\mathcal{H}\phi'+a^{2}V_{\phi}=0.
\eeq
where $\mathcal{H}$ is the conformal Hubble parameter, $\phi$ is the scalar field, $V$ represents its potential, $\kappa^{2}=8\pi G$, and $V_{\phi}=\frac{dV}{d\phi}$.

\begin{figure*}[htp]
\begin{center}
\includegraphics[width=0.6\textwidth]{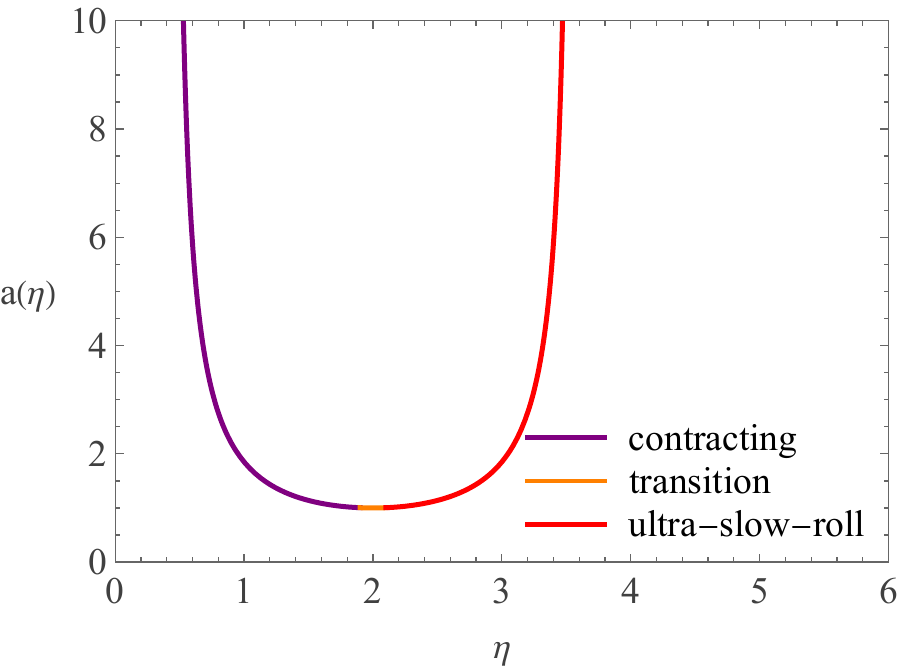}
\caption{\label{Fig1} Evolutionary curve of $a(\eta)$ for pre-inflationary bounce scenario with $\eta_{b}=2$ and $\delta\eta=0.1$.}
\end{center}
\end{figure*}

For the spatially flat universe, the primordial power spectrum of the bouncing universe is calculated by considering an instantaneous transition between a kinetically dominated contracting phase and a nearly de Sitter phase~\cite{Piao2004}. For the KICI scenario, the analytical approximation of the primordial power spectrum is studied by considering an instantaneous transition between the kinetically dominated stage and the ultra-slow-roll inflation~\cite{Thavanesan2021}. Such an instantaneous transition is unsmooth and implies an implicit inflationary potential. For the case of a pre-inflationary stage dominated by the potential term in the closed universe, we analyze the pre-inflationary bounce in a closed universe by considering a transition between a potential dominated contracting stage and an ultra-slow-roll inflationary stage. We include the transition stage here in order to analyze its impact on the primordial power spectrum and the CMB TT spectrum, together with the imprints it leaves on them. By adopting the analytical method in~\cite{Thavanesan2021, Shumaylov2022, Huang2022, Huang2023a, Huang2023b} and combining Eqs.~(\ref{CB1}) and~(\ref{CB2}), we obtain
\beq
\mathcal{H}'-\mathcal{H}^{2}-1=-\frac{1}{2}\kappa \phi'^{2} \simeq 0,
\eeq
which has the solution
\beq
a=\frac{b_{2}}{\cos(\eta+b_{1})},
\eeq
with $b_{1}$ and $b_{2}$ are integration constants. Then, the scale factor $a$ in the potential dominated contracting stage, the transition stage, and the ultra-slow-roll inflation stage can be written as
\bea
&& a=\frac{1}{\cos(\eta_{b}-\eta)}, \qquad 0<\eta<\eta_{b}-\delta\eta,\label{bua1}\\
&& a=1+\frac{1}{2}(\eta-\eta_{b})^{2}, \quad \eta_{b}-\delta\eta \leq \eta \leq \eta_{b}+\delta\eta,\label{bua2}\\
&& a=\frac{1}{\cos(\eta-\eta_{b})}, \qquad \eta_{b}+\delta\eta < \eta < \eta_{b}+\frac{\pi}{2},\label{bua3}
\eea
where $\eta_{b}$ is the conformal time at the bouncing point, and $2\delta\eta$, with $\delta\eta$ denoting a small time interval, is the duration of the transition stage. An example of the evolution of the scale factor $a(\eta)$ is shown in Fig.~\ref{Fig1}.

\section{Primordial power spectra and CMB TT spectra}

\subsection{primordial power spectra}

To calculate the primordial power spectrum for pre-inflationary bounce scenario, we need to solve the Mukhanov--Sasaki equation for curved spacetime, expressed in momentum space as~\cite{Handley2019, Thavanesan2021}
\beq\label{CMS}
v_{k}''+\Big[\mathcal{K}^{2}-\Big(\frac{\mathcal{Z}''}{\mathcal{Z}}+2K+\frac{2K\mathcal{Z}'}{\mathcal{HZ}}\Big)\Big]v_{k}=0,
\eeq
with
\beq
\mathcal{Z}=\frac{a \phi'}{\mathcal{H}}\sqrt{\frac{\mathcal{D}^{2}}{\mathcal{D}^{2}-K\mathcal{\varepsilon}}}, \qquad \mathcal{\varepsilon}=\frac{\phi'^{2}}{2\mathcal{H}^{2}}, \qquad \mathcal{D}^{2}=-\mathcal{K}^{2}+3K.
\eeq
In a closed universe, $\mathcal{K}^{2}$ is given by $\mathcal{K}^{2}=k(k+2)$, where $k$ is an integer satisfying $k>2$. After solving Eq.~(\ref{CMS}) for the Mukhanov variables $v_{k}$, the curved primordial power spectrum of the comoving curvature perturbation $\mathcal{R}$ is derived as follows
\beq\label{CPPS}
\mathcal{P_{R}}=\frac{k^{3}}{2\pi^{2}} \left| \mathcal{R}_{k} \right|^{2}=\frac{k^{3}}{2\pi^{2}} \left| \frac{v_{k}}{\mathcal{Z}_{k}} \right|^{2}.
\eeq

For the contracting stage, using Eqs.~(\ref{CB1}),~(\ref{CB2}), and~(\ref{bua1}), the relevant terms in the Mukhanov--Sasaki equation~(\ref{CMS}) can be written as
\beq
\frac{\mathcal{Z}''}{\mathcal{Z}}+2+\frac{2\mathcal{Z}'}{\mathcal{HZ}}=3.
\eeq
Then, we can write the Mukhanov--Sasaki equation for the contracting stage as
\beq
v_{k}''+k_{c}^{2}v_{k}=0, \qquad k_{c}^{2}=k(k+2)-3,
\eeq
which has the solution
\beq\label{buvk1}
v_{k}=\frac{1}{\sqrt{2k_{c}}}e^{-i k_{c} \eta},
\eeq
where the Bunch--Davies vacuum condition is taken into consideration. 

For the transition stage, we obtain
\beq
\frac{\mathcal{Z}''}{\mathcal{Z}}+2+\frac{2\mathcal{Z}'}{\mathcal{HZ}} \approx \frac{53k(k+2)-34}{10k(k+2)-5},
\eeq
from which the Mukhanov--Sasaki equation is derived as
\beq
v_{k}''+k_{b}^{2}v_{k}=0, \qquad k_{b}^{2}=k(k+2)-\frac{53}{10}+\frac{3}{4k(k+2)-2},
\eeq
whose solution is
\beq\label{buvk2}
v_{k}=\frac{k_{b}-k_{c}}{2\sqrt{2k_{c}}k_{b}}e^{-i(k_{b}+k_{c})(\eta_{b}-\delta\eta)}e^{ik_{b}\eta}+\frac{k_{b}+k_{c}}{2\sqrt{2k_{c}}k_{b}}e^{i(k_{b}-k_{c})(\eta_{b}-\delta\eta)}e^{-ik_{b}\eta}.
\eeq
Here, Eq.~(\ref{buvk1}) and the continuity conditions for $v_{k}$ and $v_{k}'$ at $\eta_{b}-\delta\eta$ are used.

\begin{figure*}[htp]
\begin{center}
\includegraphics[width=0.6\textwidth]{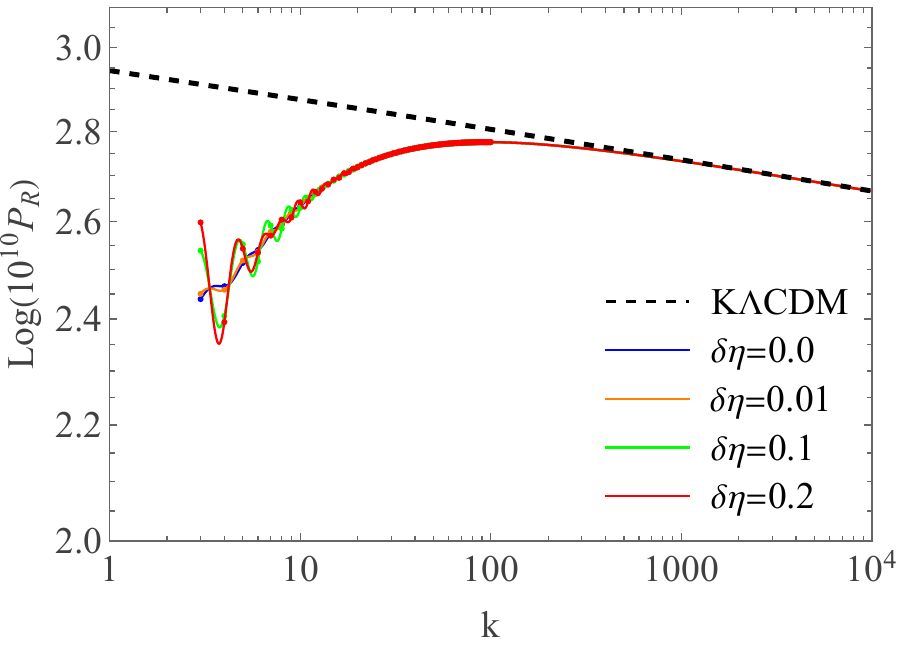}
\caption{\label{Fig2} Primordial power spectrum for the pre-inflationary bounce scenario with and without a transition stage ($\delta\eta=0.0$).}
\end{center}
\end{figure*}

For the ultra-slow-roll stage, using the limit $\varepsilon \rightarrow 0$, we find~\cite{Thavanesan2021}
\beq
\frac{\mathcal{Z}''}{\mathcal{Z}}+2+\frac{2\mathcal{Z}'}{\mathcal{HZ}} \approx \frac{2}{[(\eta_{b}+\frac{\pi}{2})-\eta]^{2}}+\frac{8}{3},
\eeq
yielding the Mukhanov--Sasaki equation
\beq
v_{k}''+\Big[ k_{i}^{2}-\frac{2}{((\eta_{b}+\frac{\pi}{2})-\eta)^{2}} \Big]v_{k}=0, \qquad k_{i}^{2}=k(k+2)-\frac{8}{3},
\eeq
with the solution
\beq\label{buvk3}
v_{k}=\sqrt{\frac{\pi}{4}}\sqrt{(\eta_{b}+\frac{\pi}{2})-\eta}\Big[C_{k}H^{(1)}_{3/2}\Big(k_{i}\Big((\eta_{b}+\frac{\pi}{2})-\eta\Big)\Big)+D_{k}H^{(2)}_{3/2}\Big(k_{i}\Big((\eta_{b}+\frac{\pi}{2})-\eta\Big)\Big)\Big],
\eeq
where $C_{k}$ and $D_{k}$ are the coefficients of the two modes of the Mukhanov variable $v_{k}$ in the ultra-slow-roll stage, which are determined by combining Eqs.~(\ref{buvk2}) and ~(\ref{buvk3}), and using the continuity conditions for $v_{k}$ and $v_{k}'$ at $\eta_{b}+\delta\eta$, and are given by
\bea
&& C_{k}=\alpha\Big[\beta H^{(2)}_{1/2}\Big(\frac{1}{2}k_{i}(\pi-2\delta\eta)\Big)-\gamma H^{(2)}_{3/2}\Big(\frac{1}{2}k_{i}(\pi-2\delta\eta)\Big)\Big],\\
&& D_{k}=-\alpha\Big[\beta H^{(1)}_{1/2}\Big(\frac{1}{2}k_{i}(\pi-2\delta\eta)\Big)-\gamma H^{(1)}_{3/2}\Big(\frac{1}{2}k_{i}(\pi-2\delta\eta)\Big)\Big],
\eea
with
\bea
&& \alpha=\frac{\sqrt{\pi}}{16k_{b}\sqrt{k_{c}(\pi-2\delta\eta)}} e^{-i(2k_{b}\delta\eta+k_{c}(\eta_{b}-\delta\eta))},\\
&& \beta=2i k_{i}(\pi-2\delta\eta)(k_{b}+k_{c}+e^{4i k_{b}\delta\eta}(k_{b}-k_{c})),\\
&& \gamma=2e^{4i k_{b}\delta\eta}(k_{b}-k_{c})(2i+k_{b}(\pi-2\delta\eta))+2(k_{b}+k_{c})(2i-k_{b}(\pi-2\delta\eta)).
\eea
Then, the primordial power spectrum is expressed as by combining Eqs.~(\ref{CPPS}) and~(\ref{buvk3}) as
\beq
\mathcal{P_{R}}=A_{s} \Big(\frac{k}{k_{*}}\Big)^{n_{s}-1} \frac{k^{3}}{k_{i}^{3}} \left| C_{k}-D_{k} \right|^{2},
\eeq
which is plotted in Fig.~\ref{Fig2}. To plot this figure, we adopt the Planck 2018 results for the curved universe best-fit data(TT,TE,EE+lowl+lowE+lensing), with  $A_{s}=2.0771 \pm 0.1017 \times 10^{-9}$ and $n_{s}=0.9699 \pm 0.0090$. In this figure, we plot the primordial power spectrum for the pre-inflationary bounce scenario with different half-duration of the transition stage $\delta\eta$, where $\delta\eta=0$ corresponds to the case without a transition stage. This figure shows that the primordial power spectra for the pre-inflationary bounce scenario are suppressed, and the spectra oscillate for $k<20$, with an amplitude that increases with the half-duration of the transition stage $\delta\eta$. The primordial curvature perturbations from the pre-inflationary stage are transferred to the inflationary stage through the transition phase; the oscillation amplitude is determined by the details of the transition: the smoother the transition, the smaller the oscillation amplitude. Although the oscillation amplitudes differ depending on the transition, the resulting primordial power spectra share a common suppression trend at small wavenumbers.

\begin{figure*}[htp]
\begin{center}
\includegraphics[width=0.6\textwidth]{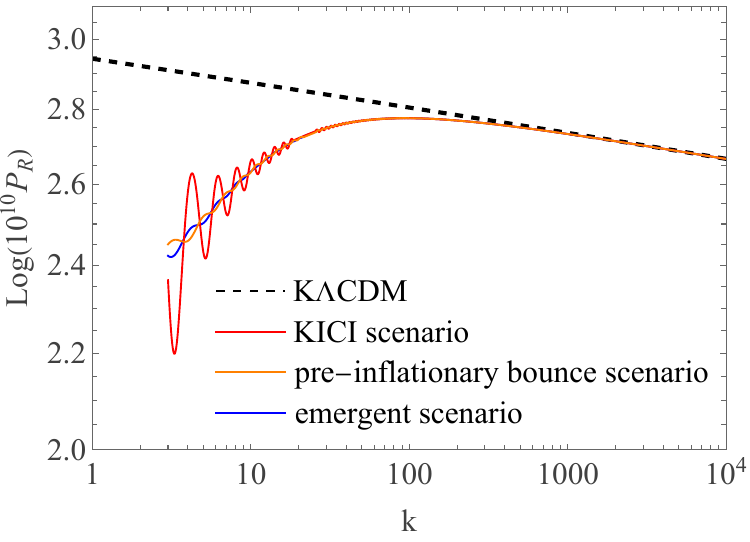}
\caption{\label{Fig3} Primordial power spectra for the KICI scenario, pre-inflationary bounce scenario, and emergent scenario in a spatially closed universe.}
\end{center}
\end{figure*}

In Fig.~\ref{Fig3}, we compare the primordial power spectra for the KICI scenario, pre-inflationary bounce scenario, and emergent scenario in the closed universe. We find that the pre-inflationary bounce scenario exhibits the same suppression trend as the KICI and emergent scenarios. This common suppression trend is present for $k<100$, while their oscillation amplitudes differ in the range $k<20$. Moreover, the analytical primordial power spectra for these models are independent of the specific form of the potential. For comparison, in the spatially flat universe, a suppression is also present in all three scenarios, although its detailed features depend on the free parameters of each model~\cite{Hergt2019a, Piao2004, Labrana2015}. This suggests that while the suppression itself arises from the pre-inflationary dynamics, the common suppression trend observed in the closed universe is instead a consequence of the spatial curvature common to all three scenarios. 

\subsection{CMB TT spectra}

\begin{figure*}[htp]
\begin{center}
\includegraphics[width=0.45\textwidth]{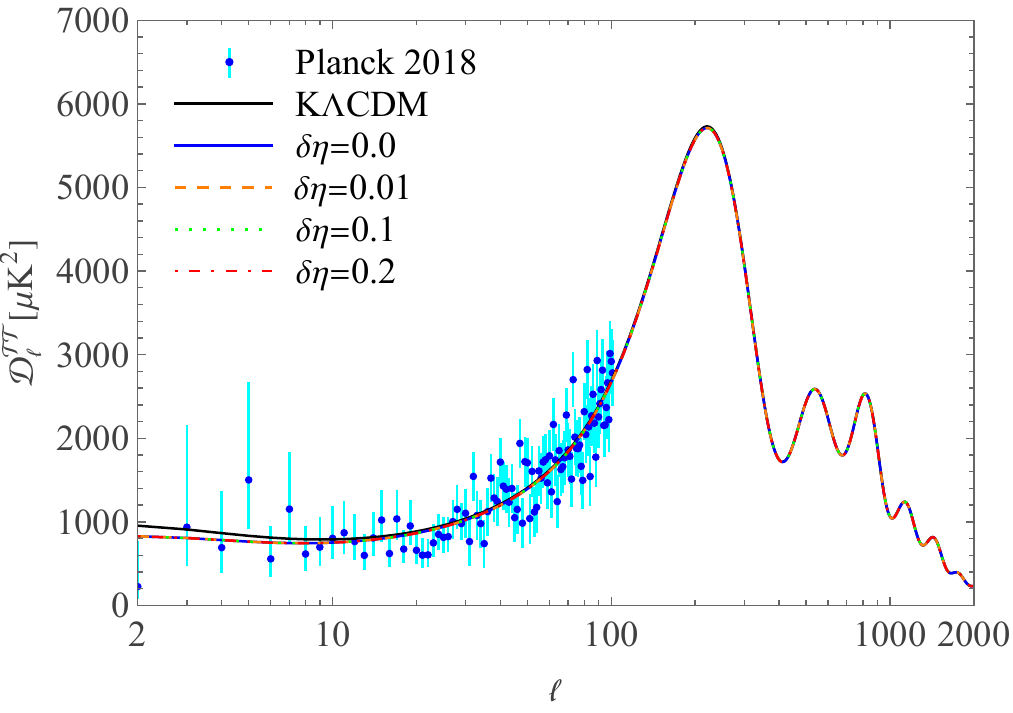}
\includegraphics[width=0.45\textwidth]{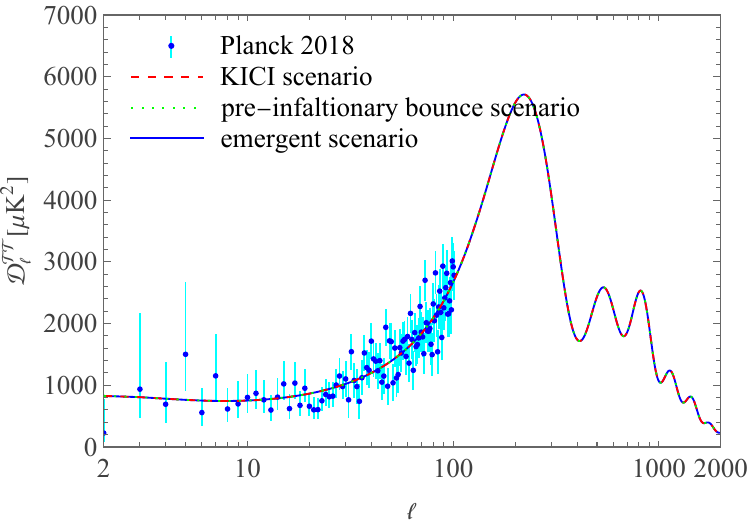}
\caption{\label{Fig4} CMB TT spectra for the KICI scenario, pre-inflationary bounce scenario, and emergent scenario in a spatially closed universe.}
\end{center}
\end{figure*}

In the previous subsection, we have analyzed the primordial power spectra for the KICI scenario, pre-inflationary bounce scenario, and emergent scenario in a spatially closed universe and found that the suppression of the primordial power spectrum for $k<100$ is present in all these scenarios. In this subsection, we discuss the suppression of the CMB TT spectrum on large scales for these scenarios. To achieve this goal, we use the CLASS code~\cite{Blas2011} to compute the CMB TT spectrum for these scenarios, as shown in Fig.~(\ref{Fig4}). 

In the left panel of Fig.~(\ref{Fig4}), we plot the CMB TT spectra for the pre-inflationary bounce scenario with different durations of the transition stage; the blue line corresponds to the spatially closed universe without a transition stage. The results show that the CMB TT spectra for the different cases exhibit a suppression for $l<10$ and overlap completely, indicating that the CMB TT spectrum is insensitive to the duration of the transition stage. In the right panel, we compare the CMB TT spectra for the KICI scenario, pre-inflationary bounce scenario, and emergent scenario. We find that the pre-inflationary bounce scenario produces the same CMB TT spectrum as the KICI and emergent scenarios, indicating that the large-scale suppression of the CMB TT spectrum is insensitive to the specific pre-inflationary dynamics and is therefore a universal feature.

The complete overlap of the CMB TT spectra originates from the low multipoles suppression ($l<10$), which arises from the common suppression trend at $k<100$ in the primordial power spectra common to all three scenarios. The differences in oscillation amplitude, which appear at $k<20$ in the primordial power spectra, are largely smoothed out by the line of sight projection in the radiation transfer, and the residual differences at low multipoles are further obscured by the large cosmic variance at these scales. Therefore, although the three scenarios have different oscillation amplitudes in their primordial power spectra, the CMB TT spectrum alone cannot distinguish among them.

To further assess the physical significance of the pre-inflationary dynamics, we compare the $\chi^{2}$ values of the CMB TT spectrum for the KICI scenario, pre-inflationary bounce scenario, and emergent scenario with that of the K$\Lambda$CDM model, and consider $2 \leq l \leq 20$ since these scenarios deviate from the K$\Lambda$CDM model for $l<20$, as summarized in Table~\ref{Tab1}. As shown in this table, the K$\Lambda$CDM model yields a worse fit to the low $l$ data with $\chi^{2}_{reduced} \simeq 2.35$, while all three scenarios reduce the $\chi^{2}$ to $\chi^{2}_{reduced} \simeq 1.75$ and produce $\Delta\chi^{2} \simeq -11.5$, which corresponds to a strong statistical preference for these scenarios over the K$\Lambda$CDM model in explaining the suppression of the CMB TT spectrum at the large scale. Meanwhile, the differences among the KICI scenario, pre-inflationary bounce scenario, and emergent scenario are less than $0.01$ in $\chi^{2}$, indicating that the CMB TT spectrum is insensitive to the specific pre-inflationary dynamics.

\begin{table}
\caption{\label{Tab1} Comparison of $\chi^2$ fits for K$\Lambda$CDM, KICI scenario, pre-inflationary bounce scenario, and emergent scenario.}
\begin{center}
\renewcommand{\arraystretch}{1.2}
 \begin{tabular}{|c|c|c|c|}
  \hline
  \hline
  Model & $\chi^{2}$ & $\chi^{2}_{reduced}$ & $\Delta \chi^{2}$\\
  \hline
  K$\Lambda$CDM & $44.6689$ & $2.3510$ & $0$\\
  \hline
  KICI scenario & $33.1570$ & $1.7451$ & $-11.5119$\\  
  \hline
  pre-inflationary bounce scenario & $33.1658$ & $1.7456$ & $-11.5031$\\
  \hline
  emergent scenario & $33.1656$ & $1.7456$ & $-11.5033$\\
  \hline
  \hline
  \end{tabular}
\end{center}
\end{table}

Thus, in the spatially closed universe, the pre-inflationary bounce scenario produces the same CMB TT spectrum as the KICI and emergent scenarios; the CMB TT spectra for the KICI, pre-inflationary bounce, and emergent scenarios are not only suppressed at large scales but also overlap completely. Moreover, since the primordial power spectra of these scenarios are independent of the specific form of the potential, the resulting CMB TT spectra inherit this potential independence. This indicates that both the suppression and the detailed shape of the CMB TT spectrum are nearly universal features, insensitive to the specific pre-inflationary dynamics and the details of the transition. The $\chi^{2}$ comparison further demonstrates that all three pre-inflationary scenarios are statistically preferred over the K$\Lambda$CDM model in explaining the observed large-scale suppression.

\section{Conclusion}

Observations show that the CMB TT spectrum is suppressed on large scales, a feature that can be accounted for by the pre-inflationary physics in the very early universe. In the spatially closed universe, it has been shown that this suppression can be realised in the KICI and emergent scenarios. In this paper, following the analytical approach of~\cite{Thavanesan2021}, we study a potential dominated pre-inflationary stage in the closed universe and construct a pre-inflationary bounce scenario that does not rely on a specific form of the potential. This is achieved by considering a transition between a potential-dominated contracting stage and an ultra-slow-roll inflationary stage.

Then, we analyze the primordial power spectrum for the pre-inflationary bounce scenario. We find that it is suppressed for $k<100$ and exhibits oscillations whose amplitude is determined by the details of the transition, with a smoother transition yielding a smaller oscillation amplitude. By comparing the primordial power spectra of the pre-inflationary bounce scenario with those of the KICI and emergent scenarios, we find that the pre-inflationary bounce scenario exhibits the same suppression trend as the KICI and emergent scenarios. This common suppression trend is present for $k<100$, while their oscillation amplitudes differ in the range $k<20$.

Based on the primordial power spectrum for the pre-inflationary bounce scenario, we then calculate the CMB TT spectrum for this scenario with different durations of the transition stage. We find that the CMB TT spectrum is insensitive to the duration of the transition stage, and that the spectra for different durations overlap completely. We further compare the CMB TT spectra for the KICI, pre-inflationary bounce, and emergent scenarios, and find that the pre-inflationary bounce scenario produces the same CMB TT spectrum as the KICI and emergent scenarios, indicating that the large scale suppression of the CMB TT spectrum is insensitive to the specific pre-inflationary dynamics and therefore represents a nearly universal feature.

To further assess the statistical significance of these CMB TT spectra, we compare the corresponding $\chi^{2}$ values for the three scenarios with that of the K$\Lambda$CDM model. The results show that the K$\Lambda$CDM model yields a worse fit to the low $l$ data than the three scenarios, and the differences among the KICI, pre-inflationary bounce, and emergent scenarios are less than $0.01$ in $\chi^{2}$. This indicates that the CMB TT spectrum is statistically insensitive to the specific pre-inflationary dynamics, and that the pre-inflationary bounce scenario can account for the large scale suppression of the CMB TT spectrum as well as the KICI and emergent scenarios.

Finally, we note that all three scenarios are analyzed within the same potential independent approach, so that their analytical primordial power spectra are independent of the specific form of the potential. This approach, however, does not predict the power spectrum amplitude $A_{s}$ and the spectral index $n_{s}$, which depend on the specific form of the potential and are therefore fixed to the Planck 2018 values. What this approach does predict is the characteristic low-$k$ shape of the primordial power spectrum, including the suppression and oscillations. The suppression of the primordial power spectrum originates from the pre-inflationary dynamics, while the common suppression trend shared by the three scenarios is naturally explained by the positive spatial curvature of the closed universe, which shifts the dynamical wavevector and thereby produces the common suppression trend in all three scenarios. This common trend is inherited by the CMB TT spectra, which overlap completely for the three scenarios.

\begin{acknowledgments}

This work was supported by the National Natural Science Foundation of China under Grants Nos.12405081, 11865018.

\end{acknowledgments}

\end{document}